\documentstyle[manuscript,aps,psfig]{revtex}

\begin{document}
\draft
\title{Output functions and fractal dimensions in
dynamical systems}
\author{Alessandro P. S. de Moura$^1$ and Celso Grebogi$^2$} 
\address{1. Institute for Plasma Research, 
University of Maryland, College Park, Maryland 20742, U.S.A.}
\address{2. Instituto de F\'{\i}sica, Universidade de S\~ao Paulo,
Caixa Postal 66318, 05315-970 S\~ao Paulo, SP, Brazil.}
\date{\today}
\maketitle

\begin{abstract}
We present a novel method for the calculation of the fractal dimension
of boundaries in dynamical systems, which is in many cases many orders
of magnitude more efficient than the uncertainty method. We call it
the Output Function Evaluation (OFE) method. The OFE method is based
on an efficient scheme for computing output functions, such as the
escape time, on a one-dimensional portion of the phase space. We show
analytically that the OFE method is much more efficient than the
uncertainty method for boundaries with $D<0.5$, where $D$ is the
dimension of the intersection of the boundary with a one-dimensional
manifold. We apply the OFE method to a scattering system, and compare
it to the uncertainty method. We use the OFE method to study the
behavior of the fractal dimension as the system's dynamics undergoes a
topological transition.
\end{abstract}


{\it Motivation. ---} In dynamical systems with two or more
well-defined asymptotic states (examples are found in systems with
attracting or repelling sets), the boundary in phase space separating
initial conditions corresponding to distinct final states (that is,
belonging to different basins) may have a fractal structure; in this
case one says that the system has a {\em fractal basin
boundary}\cite{fbb}. This means that sets of points in phase space
which undergo very different time evolutions are mixed in a complex
way in all scales. Fractal basin boundaries are found in many
important physical systems, such as in astrophysics\cite{astro},
scattering systems\cite{scattering}, systems with
escapes\cite{escapes}, dissipative systems\cite{dissip}, etc. The
occurrence of fractal basin boundaries implies a great sensitivity of
the long-time evolution of the system to perturbations in the initial
conditions. This sensitivity to initial conditions is characterized by
the {\em box-counting dimension} $d$ of the basin boundary, which is
interpreted as a measure of the degree of uncertainty about the final
fate of a system with fractal basin boundaries\cite{fdim}. The
box-counting dimension is one of the most important quantities for
characterizing these boundaries. Denoting the dimension of the total phase
space by $d_{ps}$, $d$ satisfies in general $d_{ps}-1\le d<
d_{ps}$. $d=d_{ps}-1$ for regular boundaries, and $d>d_{ps}-1$ for
fractal boundaries.

Because of its fundamental physical significance, it is very important
to have efficient numerical methods for calculating $d$ with good
precision. The best method known so far is the {\em uncertainty
method}, which is based on a direct exploitation of the final state
uncertainty reflected in $d$\cite{fdim}. The uncertainty method is
very efficient for high values of $d$ (that is, for $d$ close to
$d_{ps}$), but it is inefficient for low values of $d$ (close to
$d_{ps}-1$), when the basin boundary departs only a little from a
smooth manifold. This is because the number of initial points whose
orbits are integrated for a calculation of $d$ in this method for a
one-dimensional set of initial conditions behaves as $\epsilon^{D-1}$,
where $D$ is the {\em reduced dimension} $D=d-d_{ps}+1$, and
$\epsilon$ is the smallest scale used in the computation, which
defines the precision in the calculation: the smaller $\epsilon$ is,
the higher the precision will be. $D$ is the dimension of the
intersection of the boundary with a generic one-dimensional segment in
phase space, and is bounded by $0\le D\le 1$. We see that for $D$
close to 0, this method is very inefficient, because in this case the
number of integrated orbits grows rapidly as $\epsilon$
decreases. Motivated by this, we introduce in this Letter a new method
for the numerical calculation of the box-counting dimension that is
highly efficient for small values of $D$, and it is in this respect
complementary to the uncertainty method. We call our method the {\em
Output Function Evaluation} (OFE) method. Specifically, we show that
the number of integrated orbits in the OFE method scales as
$\epsilon^{-D}$. We then apply the method to a scattering system and
compare its results and performance with the uncertainty method. We
show that the two methods give the same result (as of course they
should), but that for low $D$ our method is several orders of
magnitude more efficient. Finally, we apply the OFE method to this
scattering system and we show that the fractal dimension shows a
characteristic behavior at a critical energy for which the invariant
set suffers a topological transition.

{\it Method. ---} We start with a brief exposition of the uncertainty
method. Consider a one-dimensional set $C$ in the phase space. Take
now a pair of points in $C$ separated by a distance $l$, in a random
position in $C$. The probability $P(l)$ that the two points belong to
different basins scales as $P(l)\sim l^{1-D}$. The uncertainty method
amounts to a direct calculation of $P(l)$ for many values of $l$,
thereby obtaining $D$. This is done by choosing randomly many pairs of
$l$-separated points, and numerically integrating their corresponding
trajectories. For a large number of pairs, the fraction of pairs
which evolve to distinct final states for a given separation $l$
should approach $P(l)$. Doing this for several values of $l$, one can
find $D$ by fitting $P(l)$ to a power law. The precision of the
resulting $D$ obtained in this way depends ultimately on the smallest
value for $l$ used in the computation, which we denote by
$\epsilon$. For $\epsilon$ small enough, the total number of initial
conditions $N_{unc}$ integrated in the computation of the uncertainty
method scales as
$1/P(\epsilon)$, that is,
\begin{equation}
N_{unc}(\epsilon) \sim \epsilon^{D-1}.
\label{nunc}
\end{equation}
Since the computation time is roughly proportional to $N_{unc}$, one sees
that although being very efficient for $D$ close to one, this method
is definitely 
inefficient for $D$ near zero.

The reason why the uncertainty method is inefficient for $D$ close to
zero is that for low $D$ the volume occupied by the boundary for a
given resolution $\epsilon$ (with $\epsilon$ small) is very small, and
the vast majority of pairs of initial conditions will fall on the
same basin. Since one needs to find a minimum number of pairs of
points belonging to different basins in order to have a reasonable
statistics for $P$, this causes $N_{unc}$ to become very large, and to
grow very rapidly as $\epsilon$ decreases. To calculate low $D$ values, we
need a method that ``focuses its attention'' on the vicinity of the
boundary, and concentrates its evaluations there. This is exactly what
the OFE method does, and we explain it next.

Our method is based on the computation of suitable output functions of the
system; these are functions relating the values of particular
variables of the system after it reaches one of the final states
(e.g., after it converges to a neighborhood of an attractor in a
dissipative system) to the initial conditions
that led to that time evolution. One example is the time $\tau$ it
takes to reach the final state; another example is the deflection
angle $\phi(b)$ as a function of the impact parameter $b$ in
scattering systems. We restrict ourselves to output functions whose
domain is a one-dimensional sub-manifold of the phase-space which
intersects the basin boundary. If the boundary is fractal, so is its
intersection with this sub-manifold. The output functions mirror the geometrical
structure of the basin boundary, and if the boundary is fractal, so
are the output functions, and in this case they have a fractal set of
singularities with the same dimension $D$ as the basin boundary. 

Although our method works with many choices of output functions, in order to
clarify the exposition, from now on we assume for definiteness that
we are dealing with a scattering system, and we choose the output
function to be the deflection angle $\phi(b)$. We have to calculate
$\phi$ with a resolution given by $\epsilon$, with $\epsilon$ being
very small. The straightforward method of laying on a given interval of
$b$ an $\epsilon$-size grid and calculating $\phi$ for the
points on the grid is not good, since the number of integrations goes
as $\epsilon^{-1}$, which is even more inefficient than the
uncertainty method. To improve this, we use a variable-sized grid. The
idea is to adjust the size of the grid on $b$ (the
{\em stepsize}) so that the oscillations in $\phi$ are well resolved, with
the minimum size of the grid given by the resolution $\epsilon$. For
values of $b$ away from the boundary, $\phi$ is smooth, and the grid
size can be large, whereas for $b$ close to the boundary, $\phi$ is
very steep and typically shows very wild oscillations; in this latter
case, the grid size needs to be small to resolve $\phi$. The steepness
of $\phi$ for a given $b$ is measured by the modulus of its derivative
$|d\phi/db|$, and we choose the gridsize $\Delta$ to be proportional
to $1/|d\phi/db|$, with the constraint $\Delta\ge\epsilon$. In this
way, most of the computing will happen near the fractal region of
$\phi$, that is, near the basin boundary.

The method is implemented as follows. For a given $b$ interval
$(b_{in},b_{fin})$, $\phi$ is calculated sequentially for a set of $b$ values
$b_0,b_1,b_2,\cdots,b_j,\cdots$, with $b_j<b_{j+1}$. We proceed by
first choosing $b_0=b_{in}$, $b_1=b_{in}+\epsilon$, and by integrating
initial conditions corresponding to $b_0$ and $b_1$, we compute
$\phi(b_0)$ and $\phi(b_1)$, which we denote by $\phi_0$ and $\phi_1$,
respectively. Generally, we use the notation
$\phi_j\equiv\phi(b_j)$. Now from $b_0$ and $b_1$ we obtain $b_2$ from
$b_2=b_1+\Delta_1$, where the stepsize $\Delta_j$ is given by
\begin{equation}
\Delta_j = \left\{
\begin{array}{lr}
\xi_j, & \mbox{if $\epsilon\le\xi_j\le\Delta_{max}$} \\
\epsilon, & \mbox {if $\xi_j<\epsilon$} \\
\Delta_{max}, & \mbox{if $\xi_j>\Delta_{max}$} \\
\end{array}
\right.
\label{stepsize}	
\end{equation}	
with
\begin{equation}
\xi_j = \min\left(\frac{\delta}{|d\phi(b_j)/db|},\alpha\Delta_{j-1}\right),
\label{eqxi}
\end{equation}
where $\Delta_{max}$, $\delta$, and $\alpha$ are constant parameters.
$d\phi(b_j)/db$ is the derivative of $\phi$ calculated at $b=b_j$. The
idea is that the stepsize $\Delta_j$ be chosen so that
$\phi_{j+1}-\phi_j\approx\delta$, to a first-order approximation. In
other words, the stepsize is chosen so that the variation of $\phi$
from one point to the next is kept approximately constant; this is the
key idea of our method. However, we
do not allow the stepsize to grow too much from one step to the next:
Eq. (\ref{eqxi}) ensures that $\Delta_{j}/\Delta_{j-1}\le\alpha$, with
$\alpha>1$ giving the constraint on the growth of the stepsize. This
avoids problems near extrema, where $d\phi(b)/db=0$ and the
first-order estimate of $\phi_{j+1}-\phi_j$ is not valid. Also,
$\Delta_j$ is restricted to be within the interval
$(\epsilon,\Delta_{max})$. We use the
two-point approximation for the derivative $d\phi(b_j)/db$:
\begin{equation}
\frac{d\phi(b_j)}{db} \approx \frac{\phi_j-\phi_{j-1}}{\Delta_{j-1}}.
\end{equation}
Now from $b_2$, we calculate $\phi_2$, and obtain $b_3$ through
$b_3=b_2+\Delta_2$, and so on. The computation is stopped when we
reach step $N$ for which $b_N>b_{fin}$. 

Once we have calculated $\phi(b)$ by this procedure, the fractal set of
singularities is given (to resolution $\epsilon$) by points $b_j$
such that $|b_j-b_{j-1}|>\beta$, with $\beta$
being a parameter satisfying
$\beta>\delta$. This means that we pick the points $b_j$ corresponding
to regions where the output function shows oscillations on scales
smaller than $\epsilon$. The results of the method are independent of
the value chosen for $\beta$.

Now that we are in the possession of a set of points $M$
approximating the boundary to resolution $\epsilon$, we proceed to
calculate the fractal dimension. We could use a direct
implementation of the definition of the box-counting dimension, but we
prefer to use a more powerful method, described in \cite{metd,tel}, which
we explain briefly now. For each point $b_i$ in $M$, we count the
number $n_i(l)$ of points in $M$ that lie within a distance $l$ of
$b_i$. It can then be shown\cite{metd} that the average of $1/n_i(l)$ over
all points in $M$ scales with $l$ as
\begin{equation}
\left<\frac{1}{n(l)}\right> \sim l^{-D},
\label{metodo}
\end{equation}
where $D$ is the reduced dimension. From Eq. (\ref{metodo}), we
obtain $D$ by calculating $\left<1/n(l)\right>$ for many values of $l$
(with $l$ small), 
and fitting the results to a power law.

We now estimate how the number of integrations $N_{OFE}$ of the OFE
method scales with the resolution $\epsilon$. By construction, in the
OFE method most of the integrations are performed for points in the
vicinity of singularities in the output function, where the function
is the steepest. For a small enough $\epsilon$, $N_{OFE}$ is therefore
proportional to the number of singularities that are resolved with
resolution $\epsilon$; but by the definition of the box-counting
dimension, this number is proportional to $\epsilon^{-D}$. Therefore,
we have the result
\begin{equation}
N_{OFE} \sim \epsilon^{-D}.
\label{nofe}
\end{equation}
For $D$ close to zero, $N_{OFE}$ grows very slowly, and the OFE method
is much more efficient than the uncertainty method; the opposite
holds for $D$ close to one. To better compare the two
methods, we define $f(\epsilon)$ to be the ratio of $N_{OFE}$ and
$N_{unc}$. From Eqs. (\ref{nunc}) and (\ref{nofe}), we have:
\begin{equation}
f(\epsilon) = \frac{N_{OFE}}{N_{unc}} \sim \epsilon^{1-2D}.
\label{eqf}
\end{equation}
We see that $f\rightarrow 0$ for $\epsilon\rightarrow 0$ if
$D<0.5$. This means that for $D<0.5$ (and $\epsilon$ small enough) the
OFE method is more efficient than the uncertainty method, and it
becomes ever more so as $\epsilon$ decreases. In fact, we will see in
the example that follows that the difference in efficiency can be of
many orders of magnitude. On the other hand, if $D>0.5$,
$f\rightarrow\infty$ for $\epsilon\rightarrow 0$; in this case, the
uncertainty method is the more efficient one.

{\it Example. ---} We exemplify our method with a Hamiltonian
scattering system with two degrees of freedom, described by a
potential function $V(x,y)$, where $V$ is required to be highly
localized around the origin. To exemplify our results, we use a
potential that is a superposition of three repulsive Gaussian hills:
\begin{equation}
V(x,y) = \sum_{i=1}^3V_i\exp\left(-r_i^2/2\sigma_i^2\right),
\label{pot}
\end{equation}
where
\begin{equation}
r_i^2 = (x-x_i)^2 + (y-y_i)^2.
\end{equation}
$(x_i,y_i)$ are the coordinates of the centers of the three
hills. $V_i$ and $\sigma_i$ are constants, and give the height and the
spread of each hill, respectively. Potentials of the form (\ref{pot})
are paradigms of chaotic scattering, and have been extensively
studied\cite{hills}. For our example, we choose the parameters to be:
$x_1=-x_2=4$, $y_1=y_2=0$, $x_3=0$, $y_3=2$; $V_1=V_2=10$, $V_3=1$;
$\sigma_1=\sigma_2=0.75$, $\sigma_3=0.325$. For this potential, there is
a transition from regular to chaotic scattering as the energy of the
incoming particle drops below a critical energy $E_c$, with
$E_c>1$.

We now proceed to apply the OFE method to this system. For the output
function we choose the deflection function $\phi(b)$, calculated for a
one-dimensional set of initial conditions on the segment $y=-10$,
$0<x<4$, with initial velocity parallel to the $y$ axis. The initial
position is sufficiently far away from the origin so that the particle
can be considered to be initially in free motion, and the velocity is
fixed by the energy constraint $v_0=\sqrt{2E}$. In this case, the
impact parameter $b$ is simply the $x$ coordinate. Each initial
condition is integrated until its distance from the origin becomes
greater than 10, when it can be considered to be in free motion
again. The deflection angle $\phi$ is found by integrating the
quadrature $\dot{\theta}=(xv_y-yv_x)/\sqrt{x^2+y^2}$ along the
trajectory, with $\phi$ being given by the value of $\theta$ after the
particle is scattered.
 
We now calculate $\phi(b)$ by the OFE method, with
$\epsilon=10^{-10}$, $\alpha=2$, $\delta=0.03$, and
$\Delta_{max}=10^{-3}$, for $E=0.95$. We then find the approximation
$M$ to the fractal set of singularities of $\phi$ as explained above,
using $\beta=0.5$. Next we calculate the fractal dimension from $M$,
using the method described above\cite{metd,tel}. The result is shown
in Fig. 1a, which is a plot of the logarithm of $\left<1/n(l)\right>$
as a function of $l$. The points clearly define a straight line, and
from its inclination we get the dimension: $D=0.238\pm 0.002$. We have
also calculated $D$ by the uncertainty method. In Fig. 1b, the
fraction $g(l)$ of $l$-separated pairs of initial conditions with
$\phi$ differing by at least $\beta=0.5$ are plotted against $l$, in a
log-log plot. For each $l$, we keep integrating pairs of points until
100 pairs have been found with $\phi$ differing by $\beta$ or
more. $D$ is found from a least-square fit of the points in Fig. 1b
and using Eq. (\ref{nunc}) to be $D=0.24\pm 0.02$. The results of the
two methods agree, as they should, but the result of the OFE method is
ten times more accurate than that of the uncertainty method, even
though the number of integrated points in the OFE calculation was only
about 26000, compared to about $1.5\times 10^6$ integrations that were
necessary in the uncertainty method. Note that in Fig. 2b the function
$g$ was not extended to $\epsilon=10^{-10}$, because that would
require an unreasonable number of integrations. From Eq. (\ref{nunc})
we can estimate the number of integrations $N_{unc}$ necessary to
extend $g$ down to $l=\epsilon$; using the value $0.24$ for $D$, we
find $N_{unc}$ to be over $10^{10}$, which is prohibitively large, and
is six orders of magnitude larger than the number of integrations in
the OFE method. This shows the superiority of the OFE method over
the uncertainty method for $D<0.5$.

The potential (\ref{pot}) undergoes a topological change in the phase
space as the energy drops below $E=1$, due to the appearance of a new
forbidden region around hill 3. At this energy there is a sudden
change (a {\em metamorphosis}) in the topological structure of the
invariant set\cite{hills}. We expect this metamorphosis to imply a
characteristic behavior of $D$ in the vicinity of $E=1$. To test this,
we apply the OFE method to obtain $D$ as a function of $E$ for $E$
close to $1$. The result is shown in Fig. 2. We see that $D$ has  a
minimum at $E=1$, and $D(E)$ exhibits a cusp
at this energy. Notice that a high accuracy in the computation
was necessary to resolve the behavior of $D$ at the metamorphosis. The
calculation of Fig. 3 using the uncertainty method would require a
prohibitively large amount of computation to come up with the same results.

We note that in the particular case of scattering in two dimensions,
it can be shown that all the thermodynamical quantities associated
with the fractal invariant set can be obtained from the time delay
function\cite{thermfunc}. This means that we can use the OFE method to compute any
such quantity, including for example the topological entropy.

In summary, we have presented the OFE method for the calculation of
the fractal dimension which is much more efficient than the
uncertainty method for $D<0.5$. We illustrated the method with a
scattering system, and we have shown for that case that our method is
many orders of magnitude more efficient than the uncertainty
method. We used the OFE method to show that the fractal dimension
displays a characteristic behavior at a topological transition of the
well-known three-hill system, which would have been very difficult to
be resolved by the uncertainty method.

This work was sponsored by FAPESP and ONR(Physics).

\begin{figure}
\begin{center}
\psfig{figure=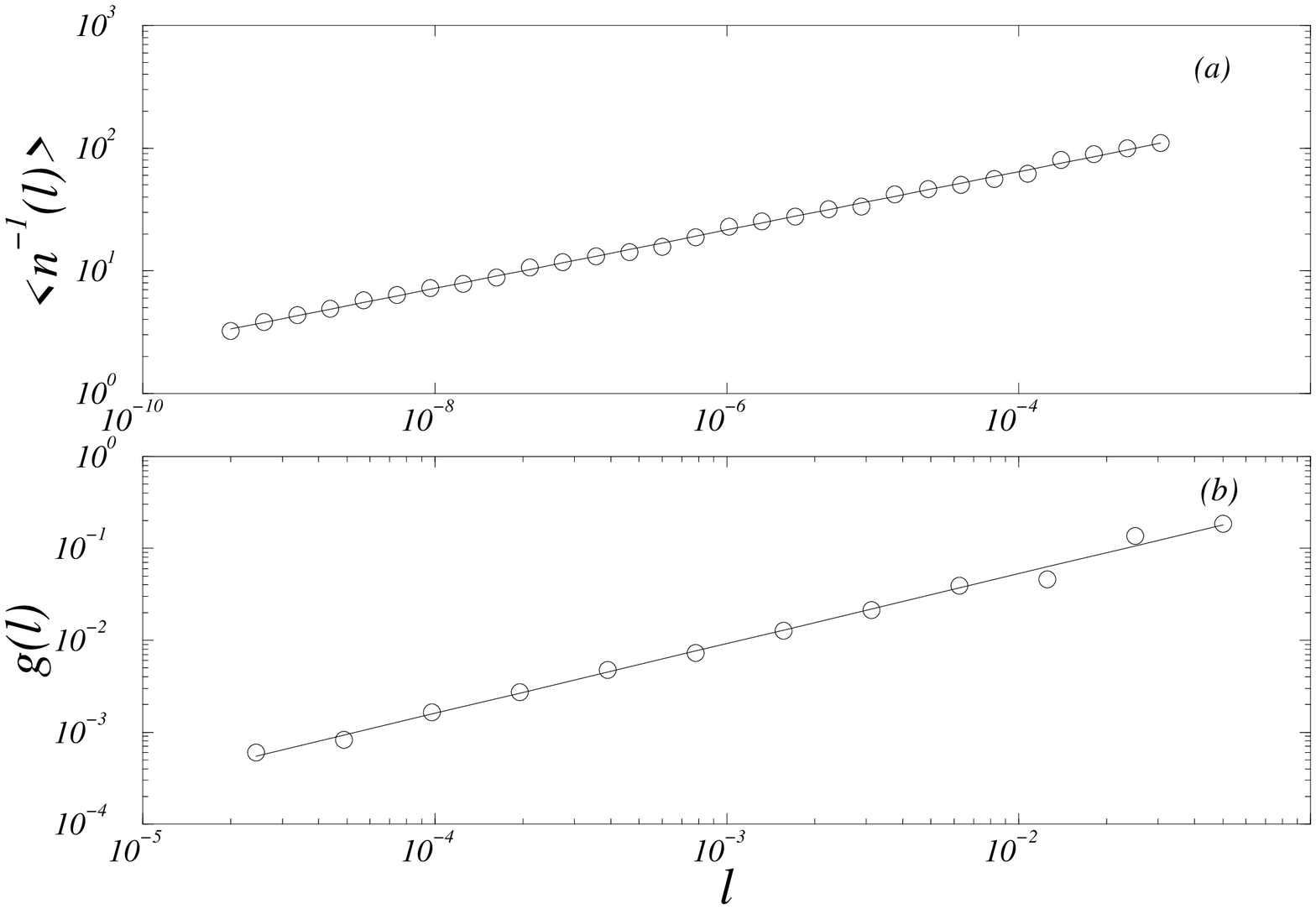,width=500pt,angle=-90}
\vspace{0.5cm}
\caption{(a) Determination of the fractal dimension by the OFE
method. The fit gives $D=0.238\pm 0.002$. (b) Result of the
uncertainty method. The fit gives $D=0.24\pm 0.02$.}
\end{center}
\end{figure}

\begin{figure}
\begin{center}
\psfig{figure=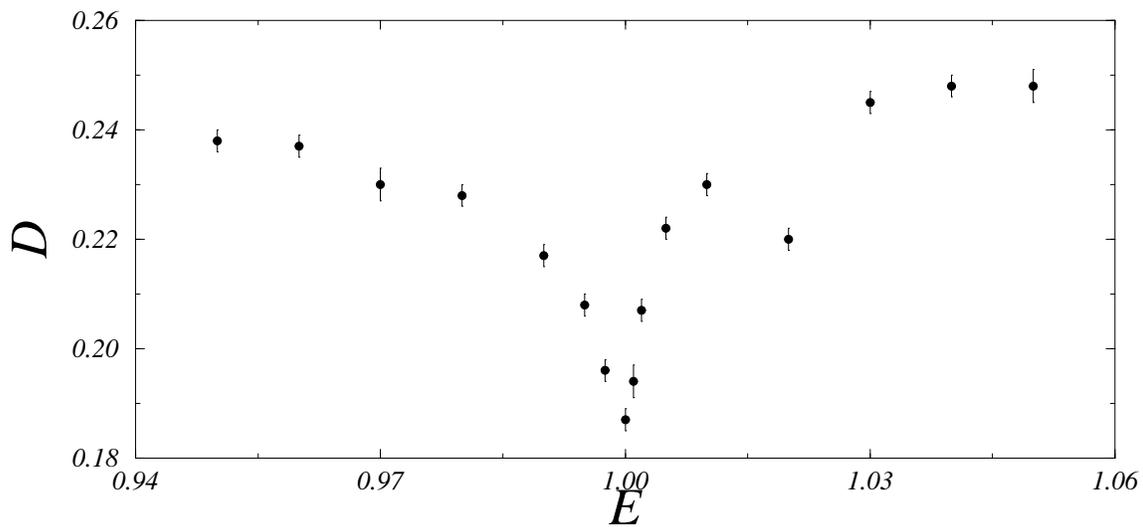,width=500pt,angle=-90}
\vspace{0.5cm}
\caption{Reduced dimension $D$ as a function of the energy $E$.}
\end{center}
\end{figure}

\end{document}